\documentclass[11pt,a4paper]{JHEP3}
\usepackage{amsmath,epsf,amssymb,latexsym,cite,graphics,bbm}
\usepackage[matrix,arrow,frame,import,curve,color]{xy}

\DeclareFontFamily{U}{rsf}{}
\DeclareFontShape{U}{rsf}{m}{n}{
  <5> <6> rsfs5 <7> <8> <9> rsfs7 <10-> rsfs10}{}
\DeclareMathAlphabet\Scr{U}{rsf}{m}{n}

\def\C{{\mathbb C}}

\def\codim{\operatorname{codim}}

\def\rank{\operatorname{rank}}
\def\GU{\operatorname{U{}}}

\def\p{\partial}
\def\pb{\bar{\partial}}

\def\la{\langle}
\def\ra{\rangle}

\def\ff#1#2{{\textstyle\frac{#1}{#2}}}

\def\cA{{\cal A}}

\def\cD{{\cal D}}
\def\cE{{\cal E}}

\def\cJ{{\cal J}}

\def\cL{{\cal L}}

\def\cN{{\cal N}}
\def\cO{{\cal O}}

\def\cQ{{\cal Q}}

\def\cW{{\cal W}}
\def\cX{{\cal X}}

\def\ep{{\epsilon}}

\newcommand\alphab{\overline{\alpha}}

\newcommand\gammab{\overline{\gamma}}

\newcommand\thetab{\overline{\theta}}

\newcommand\phib{\overline{\phi}}

\newcommand\psib{\overline{\psi}}

\newcommand\etat{\widetilde{\eta}}

\newcommand\lambdat{\widetilde{\lambda}}

\newcommand\omegat{\widetilde{\omega}}

\newcommand\Gammab{\overline{\Gamma}}

\newcommand\Phib{\overline{\Phi}}

\newcommand\Gammat{\widetilde{\Gamma}}

\newcommand\cb{\overline{c}}

\newcommand\hb{\overline{h}}

\newcommand\jb{\overline{\jmath}}

\newcommand\wb{\overline{w}}

\newcommand\zb{\overline{z}}

\newcommand\Gb{\overline{G}}

\newcommand\Jb{\overline{J}}
\newcommand\Kb{\overline{K}}

\newcommand\Ft{\widetilde{F}}

\newcommand\Rt{\widetilde{R}}

\newcommand\Vt{\widetilde{V}}

\def\cDb{\overline{\cD}}
\def\cQb{\overline{\cQ}}
\def\ACO#1#2{ \{#1,#2\} }
\def\GUB{ \GU(1)_{\text{B}}}
\def\GUL{ \GU(1)_{\text{L}}}
\def\GUR{ \GU(1)_{\text{R}}}

\title{(0,2) Landau-Ginzburg Models and Residues}
\author{Ilarion V.~Melnikov\\
\normalsize Max-Planck-Institut f\"ur Gravitationsphysik (Albert-Einstein-Institut),\\
\normalsize Am M\"uhlenberg 1, D-14476 Golm, Germany
}

\abstract{ We study the topological heterotic ring in  (0,2) Landau-Ginzburg models without a (2,2) locus.  The ring elements correspond to elements of the Koszul cohomology groups associated to a zero-dimensional ideal in a polynomial ring, and the computation of half-twisted genus zero correlators reduces to a map from the first non-trivial Koszul cohomology group to complex numbers.  This map is a generalization of the local Grothendieck residue.  The results may be applied to computations of Yukawa couplings in a heterotic compactification at a Landau-Ginzburg point.}

\preprint{AEI-2009-019}
%,PACS:  11.25.Hf, 11.25.Mj}

\begin{document}

\section{Introduction}
This is a study of topological rings~\cite{Adams:2003zy,Katz:2004nn,Adams:2005tc} in two-dimensional (0,2)-supersymmetric Landau-Ginzburg (LG) theories without a (2,2) locus.  Besides providing hands-on examples of these rings in (0,2) superconformal field theories (SCFTs), there are several reasons for undertaking this exercise.

First, while progress has been made in developing general techniques for computing similar quantities in (0,2) models with a (2,2) locus, e.g.,~\cite{Guffin:2007mp,McOrist:2008ji}, results for genuine (0,2) theories are limited to exactly soluble models, e.g.,~\cite{Blumenhagen:1995ew}, and large radius limits of non-linear sigma models, e.g.,~\cite{Bouchard:2006dn,Donagi:2006yf}.  Second, experience with (2,2) theories suggests that tools developed for (0,2) LG theories should have applications to (0,2) gauged linear sigma models and generalizations such as~\cite{Adams:2006kb, Guffin:2008pi, Adams:2009av}.  Third, orbifolds of (0,2) LG theories can be used to build phenomenologically interesting heterotic vacua, and our computations should yield information about  Yukawa couplings in these compactifications.  Finally, the results yield a neat generalization of the local residue familiar from algebraic geometry.

A topological heterotic ring~\cite{Adams:2005tc} is the (0,2) analogue of the A/B topological rings in theories with (2,2)supersymmetry~\cite{Witten:1991zz}.   It is a set of operators $\cO_{\alpha}$ with a non-singular OPE; the OPE defines a commutative, associative product on the $\cO_{\alpha}$:  
$\cO_{\alpha} \cO_{\beta} = \sum_\gamma c_{\alpha\beta}^{~~~\gamma} \cO_{\gamma} $ for some structure 
constants $c_{\alpha\beta}^{~~~\gamma}$.  The structure constants have an intimate relation to three-point functions in an associated half-twisted theory---a theory whose observables include the $\cO_{\alpha}$---for which the genus zero correlators 
\begin{equation*}
\la \cO_{\alpha}(z_1) \cO_{\beta}(z_2) \cO_{\gamma}(z_3) \ra = c_{\alpha\beta\gamma}
\end{equation*}
are independent of the world-sheet metric and hence are ``topological.'' The relation can be expressed in terms of the topological metric $\eta_{\alpha\beta} = \la \cO_{\alpha} \cO_{\beta} 1 \ra$:  
\begin{equation*}
c_{\alpha\beta\gamma} = \sum_{\delta} c_{\alpha\beta}^{~~~\delta}~\eta_{\delta\gamma}.
\end{equation*}

In this note we compute the three-point functions $c_{\alpha\beta\gamma}$ in (0,2) LG models.  After setting up conventions in section~\ref{s:action} and reviewing some general facts about (0,2) LG theories and their low energy limits in section~\ref{s:lowe}, we describe the topological heterotic ring of a (0,2) LG theory in section~\ref{s:tophet}.  The result is expressed in terms of a (0,2) superpotential, which amounts to a specification of a zero-dimensional quasi-homogeneous ideal 
\begin{equation*}
J = (J_1,J_2,\ldots,J_N)\quad\text{ in}\quad  R = \C[\phi_1,\phi_2,\ldots,\phi_n].
\end{equation*}
The observables $\cO_{\alpha}$ are in one-to-one correspondence with elements of the Koszul homology groups, associated
to the ideal $J$~\cite{Kawai:1994qy}.  Writing $\cE \simeq R^N$, and thinking of $(J_1,\ldots,J_N)$ as an element in $\cE^\vee$, these 
groups are defined by
\begin{equation*}
H_k(K_\bullet,J\wedge) = \{ \omega \in \wedge^k\cE~|~ J\lrcorner \omega = 0\}/\{\omega~|~ \omega = J\lrcorner \eta,~~~\eta\in \wedge^{k+1}\cE\},
\end{equation*} 
where we set $J\lrcorner \omega \equiv 0$ for $\omega\in R$.
The homology group $H_k(K_\bullet,J\wedge)$ is isomorphic to the Koszul cohomology group $H^{N-k}(K^\bullet,J\wedge)$.  The $H^k(K^\bullet,J\wedge)$ are given by
\begin{equation*}
H^k(K^\bullet,J\wedge) = \{ \omegat \in \wedge^k\cE^\vee~|~ J\wedge \omegat = 0\}/\{\omegat~|~ \omegat = J\wedge \etat,~~~\etat\in \wedge^{k-1}\cE^\vee\}.
\end{equation*}
In section~\ref{s:loc} we reduce the computation of the $c_{\alpha\beta\gamma}$ to a map
\begin{equation*}
\tau : H^n(K^\bullet,J\wedge) \to \C
\end{equation*}
given by
\begin{equation}
\label{eq:tauintintro}
\tau(\omegat) =  \int_{\C^n} d\mu ~\sum_{A_1,\ldots,A_n} ~\omegat_{A_1\cdots A_n} \Jb^{A_1}_1\cdots \Jb^{A_n}_{n} ~e^{-S},
\end{equation}
where $\Jb^A(\phib)$ is the complex conjugate of $J_A(\phi)$, $\Jb^A_i = \p \Jb^A/\p\phib^i$, 
$S = \sum_{A} J_A \Jb^A$, and the measure $d\mu$ is 
\begin{equation*}
d\mu = \frac{i}{2\pi} d\phi^1 \wedge d\phib^1\wedge \cdots \wedge \frac{i}{2\pi} d\phi^n \wedge d\phib^n.
\end{equation*}
In section~\ref{s:res}, we show that for an arbitrary (not necessarily quasi-homogeneous) zero-dimensional ideal $J\subset R$, $\tau$ is well-defined, i.e., 
\begin{equation*}
\tau(J\wedge \etat) =0\quad\text{ and}\quad \tau(f \omegat) = 0
\end{equation*}
for all $\etat\in \wedge^{n-1} \cE^\vee$, $f \in J$ and $\omegat \in H^n(K^\bullet, J\wedge)$.  We also show that the the map is independent of variations of $\Jb^a$, so that $\Jb^A$ may be deformed at will, provided that the deformation yields a convergent integral. From these properties we infer that whenever there exists a subset 
\begin{equation*}
\sigma = \{A_1,A_2,\ldots,A_n\} \subseteq \{1,2,\ldots, N\}
\end{equation*}
such that the ideal $K = (J_{A_1},\ldots, J_{A_n})$ is zero-dimensional, then $\tau$ is given by a multi-variate residue:
\begin{equation}
\label{eq:tauresintro}
\tau(\omegat) = \frac{1}{(2\pi i)^n} \int_{\Gamma_\sigma} \frac{ \omegat_{A_1\cdots A_n}~ d\phi^1\wedge\cdots\wedge d\phi^n}{J_{A_1} \cdots J_{A_n}},
\end{equation}
where $\Gamma_\sigma$ is a real cycle
\begin{equation*}
\Gamma_\sigma = \{\phi~|~ |J_A|^2 = \ep_A>0~~\text{for}~~A\in\sigma\},
\end{equation*}
oriented by 
\begin{equation*}
d \arg J_{A_1} \wedge \cdots \wedge d\arg J_{A_n} \ge 0.
\end{equation*}
When $N=n$, $\omegat_{A_1\ldots A_n} = g~\ep_{A_1\ldots A_n}$, where  $g \in R/J$ and $\ep_{A_1\ldots A_n}$ is the completely anti-symmetric tensor with $\ep_{12\cdots n} =+1$.  In this case $\tau$ reduces to the local Grothendieck residue of $g$.  We illustrate these results in a couple of  examples in section~\ref{s:examples}.

In the case of most physical interest---where an (orbifold of an) LG model describes a special locus in the moduli space of a gauged linear sigma model for a heterotic Calabi-Yau compactification---the required set $\sigma$ exists, so that the topological heterotic ring may be computed by residues via eqn.~(\ref{eq:tauresintro}).  This general fact, as well as explicit computations in examples should be of use in exploring the moduli space of heterotic compactifications.  It should be interesting to compare the residue with the large radius sheaf cohomology computations of the Yukawa couplings.

More generally, although we do not know how to express it explicitly as a multi-variate residue, $\tau$ defined by eqn.~(\ref{eq:tauintintro}) possesses all the nice properties we expect from a residue.  It would be interesting to find a place for $\tau$ in the theory of residue currents (see, e.g., \cite{Tsikh:2004rc} for a modern review) and develop algebraic techniques to compute it.  
We suspect the map may have applications to duality for ideals along the lines of~\cite{Jouanolou:2006ed} but have not explored this.

\acknowledgments
It is a pleasure to thank K.~Altmann, C.~Beasley, A.~Dickenstein, J.~McOrist, and especially E.~Materov for useful discussions.  I would like to thank the Erwin Schr\"odinger Institute and the CERN Theory Group for hospitality while some of this work was completed.

%%%%%%%%%%%%%%%%%%%%%%%%%%%%%%%%%%%%%%%%%%%%%%%
\section{Action and Notation} \label{s:action}
The actions we consider are best described in (0,2) superspace.\footnote{A discussion of relevant
superspace details may be found in, for instance,~\cite{Distler:1995mi}.}
We work in Euclidean signature, with
superspace coordinates $(z,\zb,\theta,\thetab)$.  The superspace covariant derivatives are given by
\begin{equation*}
\cD = \p_\theta + 2 \thetab \pb,~~~\cDb = -\p_{\thetab} -2 \theta \pb,
\end{equation*}
and the supercharges are
\begin{equation*}
\cQ = \p_{\theta} - 2\thetab \pb,~~~\cQb = -\p_{\thetab} + 2\theta \pb.
\end{equation*}
Here and in what follows, $\p = \p /\p z$ and $\pb = \p / \p\zb$.

The action is specified in terms of two types of chiral superfields, the bosonic multiplets $\Phi^i$ with expansion
\begin{equation*}
\Phi^i = \phi^i + \sqrt{2} \theta \psi^i + 2\theta\thetab \pb\phi^i,~~~i=1,\ldots, n,
\end{equation*}
and the fermionic multiplets $\Gamma^A$,
\begin{equation*}
\Gamma^A = \gamma^A -\sqrt{2}\theta G^A + 2\theta\thetab \pb \gamma^A,~~~A = 1,\ldots, N.
\end{equation*}
Here $\gamma^A$ is a left-moving Weyl fermion, and $G^A$ is an auxiliary field.  Note that the pair $(\Phi,\Gamma)$ has the content of a (2,2) chiral multiplet.  We also need the conjugate fields $\Phib^i, \Gammab^A$ given by
\begin{eqnarray*}
\Phib^i          & = & \phib^i - \sqrt{2}\thetab~ \psib^i - 2\theta\thetab \pb \phib^i,\\
\Gammab_A & = & \gammab_{A} - \sqrt{2}\thetab \Gb_A -2 \theta\thetab \pb \gammab_{A}. 
\end{eqnarray*}

The supersymmetric action is 
\begin{eqnarray}
\label{eq:Ssuper}
S &=& \ff{1}{4\pi} \int d^2 z \left[ \cD \cDb \left\{ - \Phib^i \p \Phi^i + \ff{1}{2}\Gammab_A \Gamma^A \right\}  \right.\nonumber\\
~ & ~& ~~~~~~~~~~~~~\left.+ \ff{1}{\sqrt{2}} \cD \left\{\Gamma^A J_A(\Phi) \right\}
-\ff{1}{\sqrt{2}} \cDb \left\{ \Gammab_A \Jb^A (\Phib)\right\} \right].
\end{eqnarray}
After integrating out the auxiliary fields, we find the component Lagrangian
\begin{eqnarray}
\label{eq:Lcomp}
4\pi \cL &=& 2 \p\phi^i \pb \phib^i + 2 \p\phib^i \pb \phi^i + 2 \psib^i \p \psi^i + 2\gammab_{A} \pb \gamma^A \nonumber\\
~   & ~& +J_A \Jb^A - \gamma^A J_{A,i} \psi^i + \gammab_{A} \Jb^A_{i} \psib^i,
\end{eqnarray}
where
\begin{equation*}
J_{A,i} = \frac{\p J_A}{\p \phi^i},~~~ \Jb^A_{i} = \frac{\p \Jb^A}{\p \phib^i}.
\end{equation*}
The non-vanishing SUSY (anti)commutators are
\begin{equation}
\label{eq:susy}
\begin{array}{lcl}
\ACO{\cQ}{\phi^i} = \sqrt{2} \psi^i, 		  &~~~~~&	\ACO{\cQb}{\phib^i} = \sqrt{2}~ \psib^i,\\
\ACO{\cQ}{\psib^i} =2\sqrt{2}\pb \phib^i, 	  &~~~~~&	 \ACO{\cQb}{\psi^i} = -2 \sqrt{2}~\pb \phi^i,\\
\ACO{\cQ}{\gamma^A} = \sqrt{2} ~\Jb^A,    &~~~~~&	 \ACO{\cQb}{\gammab_A} = - \sqrt{2} J_A.
\end{array}
\end{equation}

In this work we study theories with a global $\GUB$  symmetry, under which the superspace coordinate $\theta$ and the
superfields $\Gamma^A$ have charge $1$, while the multiplets $\Phi^i$ remain neutral.  This R-symmetry can be used to
half-twist the theory in a manner analogous to the B-twist of a (2,2) model---hence its name.  $\GUB$ forbids certain modifications of
the theory, such as additional terms $\Gamma^A\Gamma^B\Gamma^C J_{ABC}(\Phi)$ in the superpotential or 
modification of the chirality constraint $\cDb\Gamma^A=0$ to $\cDb \Gamma^A = E^A(\Phi)$.  

The interactions in this UV Lagrangian are determined by the (0,2) superpotential $\cW_J = \Gamma^A J_A(\Phi)$. 
We will take the $J_A$ to be polynomial in the fields, so that the $J_A$ may be thought of as generators of 
an ideal $J$ in the polynomial ring $R = \C[\phi_1,\phi_2,\ldots,\phi_n]$.

\paragraph{A notational note.}   
We will frequently use the summation convention; occasionally, we will need the completely anti-symmetric tensors $\ep^{i_1\cdots i_n},~ \ep^{A_1\cdots A_N}$, with $\ep^{12\cdots n} =+1$ and $\ep^{12\cdots N} = +1$; we will denote the completely anti-symmetric part of a tensor by square brackets---e.g., $T^{[AB]}= (T^{AB}-T^{BA})/2!$.

\section{The Low Energy Limit}\label{s:lowe}
In this section we review some properties of the low energy limit of the theory defined by the Lagrangian above.  Much of this is
well-known and well-described from the linear sigma model perspective~\cite{Distler:1993mk,Distler:1995mi}.  
In order to make this work self-contained, we now give a purely LG point of view on these matters.

\subsection{Preliminaries}
We begin with a standard observation: holomorphy implies that the superpotential term $\cW_J$ is not renormalized.   We expect that the properties of the fixed point are encoded in the ideal $J$, with RG flow correcting the irrelevant kinetic terms.  In particular, we 
expect unbroken (0,2) supersymmetry if and only if there is a common solution to $J_A(\phi) = 0$.

A non-empty solution set to $J_A= 0$ in $\C^n$ is either non-compact or consists of isolated points.  We  restrict attention to the latter situation, in which case we say the ideal J is zero-dimensional.\footnote{Note that an ideal $J\subset \C[\phi_1,\phi_2,\ldots,\phi_n]$ is zero-dimensional if and only if there is a non-zero polynomial in $J \cap \C[\phi_i] $ for each $i$~\cite{Cox:1998ua}.}  It suffices to study theories with a single isolated supersymmetric vacuum, since interactions between isolated vacua will vanish in the low energy limit.  Without loss of generality, 
we take this vacuum to be the origin in $\C^n$.  

It is not difficult to show that there is a mass term in the theory whenever
\begin{equation*}
\rank( \left. J_{A,i} \right|_{\phi= 0} ) = k > 0.
\end{equation*}
In this case $k\le n$ pairs $(\Phi,\Gamma)$ will obtain masses and  may be eliminated from the low energy Lagrangian by their 
equations of motion.  Since we are interested in constructing SCFTs, we may as well assume that $k=0$.  

Experience with (2,2) LG models suggests that we restrict attention to theories with an extra global symmetry $\GUL$ that commutes with the right-moving supersymmetry.  We make the charge assignments $q[\Phi^i]= q_i$ and $q[\Gamma^A]=Q_A$, which requires
the $J_A(\Phi)$ to be quasi-homogeneous functions:
\begin{equation}
J_A(\lambda^q \Phi) = \lambda^{-Q_A} J_A (\Phi).
\end{equation}
Note that the charges are not uniquely determined, since any scalar multiple $s q_i$, $s Q_A$ will work as well.

Having made these qualifications, we are ready to describe a number of constraints on the physics of the low energy limit.  We expect 
the low energy theory to be a unitary SCFT with (0,2) supersymmetry and a $\GUL$ global symmetry.  Our goal is to extract some of 
the properties of this SCFT from the ideal $J$.

\subsection{The Left-Moving Algebra}
A remarkable feature of the (0,2) LG theory is  the existence of left-moving Virasoro and $\GUL$ current algebras in the $\cQb$-cohomology of the UV theory. This algebra, first discussed in (2,2) LG theories~\cite{Witten:1993jg} and later extended to (0,2) gauged linear sigma models in~\cite{Silverstein:1994ih} has a close relation to certain free-field realizations of minimal models obtained in~\cite{Fre:1992hp}.  To describe it, we search for operators $j,T$ with classical scaling dimensions $(h,\hb)$ respectively given by $(1,0)$ and $(2,0)$, that are quadratic in the fields and $\cQb$-closed up to equations of motion.  Up to over-all normalizations and addition of $\cQb$-exact terms, we find 
\begin{equation}
T = T_0 - \ff{\alpha}{2} \p j,
\end{equation}
where $\alpha$ is an undetermined parameter, and $T_0$ and $j$ are given by
\begin{eqnarray}
\label{eq:Tj}
T_0 & = & - \sum_A \gamma^A \p \gammab_A -2 \sum_i \p\phi^i \p \phib^i,\nonumber\\
j       & = & \sum_A Q_A \gamma^A \gammab_A -2 \sum_i q_i \phi^i \p \phib^i. 
\end{eqnarray}
It is not hard to show that $T_0$ is $\cQb$-closed up to equations of motion for any $J_A$, while $j$ is $\cQb$-closed up to equations of motion precisely when the $J_A$ are quasi-homogeneous.   Since $\left\{\cQ,\cQb\right\} = -4\pb$, $T$ and $j$ are, up to $\cQb$-exact
terms, holomorphic conserved currents.  Working up to $\cQb$-exact terms, we may compute the OPEs of $\cQb$-closed
operators using free field OPEs that follow from the Lagrangian in eqn.~\ref{eq:Lcomp}~\cite{Witten:1993jg}. This yields the promised
algebra
\begin{eqnarray}
T(z) T(w) &\sim & \frac{c/2}{(z-w)^4} + \frac{2 T(w)}{(z-w)^2} + \frac{\p T(w)}{z-w}, \nonumber\\
T(z) j(w)  &\sim & \frac{\cA}{(z-w)^3} + \frac{j(w)}{(z-w)^2} + \frac{\p j(w)}{z-w},\nonumber\\
j(z) j(w)   &\sim &  \frac{r}{(z-w)^2},\nonumber\\
j(z) \phi^i(w) &\sim & \frac{q_i \phi^i(w)}{z-w},\nonumber\\
j(z) \gamma^A(w) &\sim & \frac{Q_A \gamma^A(w)}{z-w},
\end{eqnarray}
with
\begin{eqnarray}
r  &=& \sum_A Q_A^2 - \sum_i q_i^2, \nonumber\\
c      &=& 2(n-N)-3\alpha^2 r - 6 \alpha\left[\sum_A Q_A +\sum_i q_i \right], \nonumber\\
\cA  & = & \alpha r + \sum_A Q_A + \sum_i q_i.
\end{eqnarray}
Note that the charges assigned by $j$ to the fields $\phi^i$ and $\gamma_A$ are just those of the $\GUL$ symmetry.

Given this elegant structure, it is tempting to assume that the operators $T,j$ in the $\cQb$ cohomology correspond to, respectively, the energy-momentum tensor and the $\GUL$ current in the IR.  Since we expect the SCFT to be unitary and non-trivial, it must be that $r > 0$ and $\alpha\neq 0$.  Furthermore, we expect the $\GUL$ current to be non-anomalous, i.e., $\cA =0$, which only holds for a specific value of $\alpha$:
\begin{equation}
\alpha =- \frac{1}{r} \left[\sum_A Q_A + \sum_i q_i\right].
\end{equation} 
It is convenient to absorb $\alpha$ into a normalization of the $\GUL$ charges.  In this case, $\cA = 0$ is the familiar condition~\cite{Distler:1993mk,Kawai:1994qy} 
\begin{equation}
\label{eq:chargenorm}
\sum_{A} Q_A^2 - \sum_i q_i^2 = -\sum_A Q_A -\sum_i q_i.
\end{equation}
With this normalization, we find 
\begin{eqnarray}
\label{eq:crleft}
r & = & -\sum_A Q_A - \sum_i q_i, \nonumber\\
c & = & 3r + 2(n-N).
\end{eqnarray}

\subsection{The Right-Moving R-Current and Anomaly Matching}
Barring accidental continuous symmetries, the right-moving R-symmetry $\GUR$ must be a linear combination of $\GUB$ and $\GUL$, leading to 
charges $a q_i$ for $\Phi^i$ and $a Q_A+1$ for $\Gamma^A$,  for some real $a$.   

A simple argument suggests that $a = 1$.  Consider the following OPE in $\cQb$-cohomology:
\begin{equation*}
T(z) \phi^i(w) \sim \frac{q_i}{2}\frac{ \phi^i(0)}{(z-w)^2} + \frac{\p \phi^i(0)}{z-w}.
\end{equation*}
We see that $\phi^i$ corresponds to a primary field of weight $h_i =q_i/2$.  If we (reasonably) assume that $\phi^i$ is chiral-primary on the right, then its weight should be $\hb_i = a q_i /2$.   Since RG flow does not modify the spin of an operator, it must be $h_i = \hb_i$, i.e., $a=1$.

An anomaly matching computation confirms that $a=1$ and leads to additional insight~\cite{Distler:1993mk,Distler:1995mi}. In the IR we expect a theory with a holomoprhic current $j_z$ for the left-moving $\GUL$ current algebra and an anti-holomorphic current $\jb_{\zb}$ for the right-moving R-symmetry, with two-point functions 
\begin{eqnarray*}
\la j_z (z) j_z(w) \ra = R^{\text{IR}} (z-w)^{-2}, ~~~~~ \la \jb_{\zb}(z) \jb_{\zb} (w) \ra =  \Rt^{\text{IR}} (z-w)^{-2}.
\end{eqnarray*}
The right-moving $N=2$ algebra implies  $\Rt^{\text{IR}} =\cb/3$.
We can couple this theory to background gauge fields $V, \Vt$ and compute the effective action $W[V,\Vt]$ defined by
\begin{equation}
e^{-W[V,\Vt] + W[0] } = \la \exp\left[- \frac{1}{2\pi} \int d^2 z \left[ j_z V_{\zb} + \jb_{\zb} \Vt_{z} \right] - S_{\text{c.t.}}\right] \ra,
\end{equation}
where $S_{\text{c.t.}}$ is a choice of counter-terms.  
Choosing $S_{\text{c.t.}}$ to yield the symmetric form of the anomaly, the gauge variation of $W[V,\Vt]$ under gauge transformations 
$\delta V = d \lambda$ and $\delta \Vt = d \lambdat$ has the form 
\begin{equation}
\label{eq:anomIR}
-\delta W = \frac{i}{2\pi} \int \left[ \frac{R^{\text{IR}}}{2} \lambda F - \frac{\Rt^{\text{IR}}}{2} \lambdat \Ft \right],
\end{equation}
where $ F = dV$ and $\Ft = d\Vt$.  

On general grounds, this anomaly should be reproducible in the UV theory.  In the free UV theory the currents 
$(J_z,J_{\zb})$ for $\GUL$ and $(\Jb_{z}, \Jb_{\zb})$ for the proposed R-symmetry $\GUR$  may be derived from the Lagrangian.
As far as the anomaly is concerned, we only need the fermion contributions to the current-current correlators.  These are
\begin{equation}
\begin{array}{ccc}
\la J_z(z) J_z(w) \ra = R (z-w)^{-2}, 					&~~~~~~~~~~ & \la J_{\zb}(z) J_{\zb} (w) \ra = R' (\zb-\wb)^{-2}, \\
\la J_z(z) \Jb_z(w) \ra = S (z-w)^{-2}, 					&~~~~~~~~~~ & \la J_{\zb}(z) \Jb_{\zb} (w) \ra = S' (\zb-\wb)^{-2}, \\
\la \Jb_{\zb} (z) \Jb_{\zb}(w) \ra = \Rt (\zb-\wb)^{-2}, 		&~~~~~~~~~~ & \la \Jb_{z}(z) \Jb_{z} (w) \ra = \Rt' (z-w)^{-2},
\end{array}
\end{equation}
where 
\begin{equation}
\begin{array}{cclcccl}
R &=& \sum_A Q_A^2, 			&~~~~~~~~~&   R' &=& \sum_i q_i^2, \\
S &=& \sum_A Q_A(a Q_A +1),	&~~~~~~~~~&   S' &=& \sum_i q_i (a q_i -1), \\
\Rt &=& \sum_A (a Q_A+1)^2,	&~~~~~~~~~&   \Rt' &=& \sum_i (a q_i -1)^2.
\end{array}
\end{equation}
This leads to 
\begin{equation}
\label{eq:anomUV}
-\delta W = \frac{i}{2\pi} \int \left[ \frac{R-R'}{2} \lambda F - (\alpha_{\text{c.t.}}+S') \lambda \Ft + (\alpha_{\text{c.t.}}+S)\lambdat F - \frac{\Rt'-\Rt}{2} \lambdat \Ft \right],
\end{equation}
where $\alpha_{\text{c.t.}}$ is a counterterm. This cannot match the IR anomaly unless $S'=S$, in which case $\alpha_{\text{c.t.}} = -S$ eliminates the mixed terms.  Setting $S = S'$ fixes the constant $a$ to be
\begin{equation}
a = - \frac{\sum_A Q_A + \sum_i q_i}{\sum_A Q_A^2 -\sum_i q_i^2} = 1.
\end{equation}
In the last equality we used the normalization condition of eqn.~(\ref{eq:chargenorm}).  

Having determined $a$, we read off the coefficients $R^\text{IR},\Rt^\text{IR}$ in terms of the UV data:  the former yields $R^\text{IR} = r$, matching the expectations from the left-moving algebra in the $\cQb$-cohomology; the latter yields the right-moving central charge:
\begin{equation}
\frac{\cb}{3} = \Rt^{\text{IR}} = \sum_i (a q_i -1)^2-\sum_A (1+a Q_A)^2 = r + n-N.
\end{equation}
Amusingly,  anomaly matching requires $a$ to take on the value that maximizes $\cb$ as a function of $a$.

\subsection{Checks and Caveats}
A rough check on the values of $c,\cb$ computed above is obtained by comparing the UV and IR coefficients in the diffeomorphism anomaly.  In the free theory the central charges are $\cb^{\text{free}} = 3n$ and $c^{\text{free}} = 3n + (N-n)$, so that  the Diff anomaly is proportional to $c^{\text{free}}-\cb^{\text{free}} = N-n$. This is just what we find from the IR central charges.

Evidence that the structures based on $T, j$  are relevant to the low-energy limit of 
LG theories mainly comes from experience with (2,2) theories. The theory would have $(2,2)$ supersymmetry 
if $N=n$ and $J_{i} = \p W / \p \Phi^i$.  In this case, the correspondence between N=2  minimal models and (2,2) 
LG theories~\cite{Martinec:1988zu,Vafa:1988uu} has been quantitatively tested by comparing chiral 
rings~\cite{Martinec:1988zu,Lerche:1989uy}, comparing the elliptic genera computed in the two 
descriptions~\cite{Witten:1993jg,DiFrancesco:1993dg},  as well as via results on exact RG flows between the UV and IR 
descriptions, e.g.,~\cite{Fendley:1993pi,Grisaru:1994dx}.  All of these results are consistent with the computations based on
the left-moving algebra in the $\cQb$-cohomology.

Additional evidence has been obtained via the (0,2) Calabi-Yau/LG correspondence~\cite{Witten:1993yc,Distler:1993mk}.  One can, for instance, 
check the matching of elliptic genera computed in the sigma model and LG descriptions~\cite{Kawai:1994qy}.  In addition, comparisons have 
been made between certain exactly solvable (0,2) theories and LG descriptions~\cite{Blumenhagen:1995ew}.  We are not aware of any work
generalizing the exact flow computations to (0,2) theories; it would certainly be interesting to have such results.

Since we expect the low energy limit to be a unitary theory, there are some simple constraints on $r,\cb$ that must be satisfied.  For instance, we  must have $1\le \cb \le 3n$, which leads to
\begin{equation}
\frac{1}{3}+N-n \le r \le N.
\end{equation}
There are also bounds on the weights of $\cQb$-closed operators, leading to, for example, $r \ge q_i \ge 0$.
The latter is a generalization of the familiar (2,2) bound on the weights of chiral primary operators: $ c/6 \ge h \ge 0$. 

There are certainly cases where a blind application of the methods of the previous section patently gives nonsensical results.  For example, consider a theory with $n=1$, $N=2$, and a (0,2) superpotential $\cW_J$ given by
\begin{equation*}
\cW_J = \Gamma^1 \phi^{k+1} + \Gamma^2 \phi^{k+1}.
\end{equation*}
The low energy limit is completely transparent:  $\Gamma_- = \Gamma^1-\Gamma^2$ is a free left-moving Weyl fermion,
while $\Gamma_+= \Gamma^1+\Gamma^2$ and $\Phi$ combine to a (2,2) supersymmetric multiplet interacting with superpotential
$W = \Phi_{\text{(2,2)}}^{k+2}$.  Thus, the IR theory is a product CFT of a free left-moving Weyl fermion and the $A_{k+1}$
(2,2) minimal model with $\cb = 3k /(k+2)$.  Using the expressions above, we instead find
\begin{equation*}
\cb = 3 \frac{2 k^2}{2(k+1)^2-1}.
\end{equation*}
When $k=1$, the naive central charge is $\cb = 6/7$, which does not satisfy the unitarity bound $\cb \ge1$.

In the previous example an extra continuous global symmetry due to the free multiplet $\Gamma_-$ invalidated our assumptions.  More generally, such a symmetry can emerge accidentally in the IR.  Suppose $\cW_J^0$ describes some (0,2) LG model where the methods described above do accurately describe the SCFT.  Now add to the UV 
theory an extra multiplet $\Gamma^{N+1}$ and consider the superpotential
\begin{equation*}
\cW_{J} = \cW_J^0 + \Gamma^{N+1}\sum_A f^A(\Phi) J_A(\Phi),
\end{equation*}
where $f^A(\Phi)$ are some polynomials chosen to preserve the quasi-homogeneity conditions.  
Since the $J_A$ are $\cQb$-descendants, we expect that the additional coupling is irrelevant.  
Indeed, there is a field re-definition with Jacobian $1$ that shifts the additional interaction into the 
irrelevant D-terms:
\begin{equation*}
\Gammat^A = \Gamma^A + f^A(\Phi) \Gamma^{N+1},~~~A=1,\ldots, N, ~~~\Gammat^{N+1} = \Gamma^{N+1}.
\end{equation*}
We conclude that the low energy limit is a product of the original theory and a free Weyl multiplet.

We expect that, up to additional free left-moving fermions, two LG theories with superpotentials
$\cW_J$ and $\cW_I$ flow to the same fixed point whenever the $J_A$, $A=1,\ldots, N$ and $I_\alpha$, $\alpha = 1,\ldots, K$ define the same ideal in $\C[\phi_1,\ldots,\phi_n]$, that is, whenever there exist $R^{A}_\alpha(\Phi)$ and $S^\alpha_A(\Phi)$ such that
\begin{equation*}
J_A = \sum_\alpha S^\alpha_A I_\alpha,~~~\text{and}~~~ I_\alpha = \sum_A R^A_\alpha J_A.
\end{equation*}
As in (2,2) theories, there may also be field-redefinitions of the bosonic multiplets that lead to equivalent IR physics.\footnote{For instance, such 
redefinitions eliminate ``parameters'' that naively appear in the (2,2) LG theories that realize the supersymmetric ADE series.}

The preceding discussion suggests a necessary condition for the methods above to be reliable: the $J_A$ must furnish a minimal collection 
of generators for the ideal, and the continuous part of the symmetry group of $\cW_J$ must be $\GUL\times\GUR$.   In what follows, our working assumption will be that this condition is also sufficient, so that the $\cQb$-cohomology and the $T,j$ algebra in the UV theory accurately describe the corresponding structures in the chiral ring of the IR theory.

\section{The Topological Heterotic Ring}\label{s:tophet}
It is well known that in a unitary (2,2) SCFT the OPE leads to a natural ring structure on the space of chiral primary operators---the chiral
ring~\cite{Lerche:1989uy}.  The basic observation is that chiral primary operators $\cO_\alpha$ have a non-singular OPE that takes 
the form 
\begin{equation*}
\cO_\alpha(z) \cO_\beta(0) = c_{\alpha\beta}^{~~\gamma} \cO_\gamma(0) + \text{O}(z).
\end{equation*}
This is easily seen as a consequence of the bound satisfied by the weight and R-charge of a chiral operator:
$h \ge q/2$.  The $\GUR$ current algebra implies a bound $h \le c/6$ for chiral primary operators, which for compact
SCFTs implies that the chiral ring is actually finite.  Thus, the finite collection of operators $\cO_\alpha$ is endowed with a commutative,
associative product.  In the corresponding twisted theory, the three-point functions on the sphere, $\la \cO_\alpha \cO_\beta \cO_\gamma\ra$, are independent of the world-sheet metric and may be computed by methods of topological field theory.  When the SCFT is defined by an RG flow from 
a weakly-coupled Lagrangian theory, these techniques allow the twisted correlators to be determined by computations in terms of the UV degrees of 
freedom~\cite{Witten:1991zz}.  The case of (2,2) LG theories provides the simplest example of these sorts of computations~\cite{Vafa:1990mu}.
 
 (0,2) SCFTs possess a chiral ring defined by the cohomology classes of  the supercharge $\cQb$.   The OPE of $\cQb$-closed 
operators defines a rich holomorphic structure on this infinite-dimensional ring~\cite{Kapustin:2005pt,Witten:2005px}.  In many theories 
of interest for heterotic compactifications, there exists an important substructure:  the topological heterotic ring~\cite{Adams:2005tc}.  A distinguishing feature of such theories is the presence of a left-moving $\GUL$ symmetry, which allows for an additional projection within
$\cQb$-cohomology onto operators with left-moving weight $h$ and $\GUL$ charge $q$ related by $h = q/2$.\footnote{In many theories there are two possible projections on the left:  $h=\pm q/2$.  This is the analogue of the A/B topological rings of general (2,2) theories.
However, in LG theories only one projection leads to a non-trivial ring.}  The OPE of two such operators has the form
\begin{equation*}
\cO_\alpha(z) \cO_\beta(0) = \sum_{\mu} f_{\alpha\beta}^{~~\mu} \cX_\mu z^{h_\mu - q_\mu/2} + \left\{\cQb, \cdot\right\},
\end{equation*}
where the sum is over chiral ring elements $\cX_\mu$.  A key observation of~\cite{Adams:2005tc} is that the spins of the
operators $\cO_{\alpha},\cO_{\beta}$ constrain the possible $\cX_\mu$ on the right-hand side. Furthermore, in a large class of theories, unitarity bounds on the weights in terms of the $\GUL$ charge imply $h_\mu \ge q_\mu/2$ for all allowed $\cX_\mu$.  Thus, although a (0,2) SCFT will typically have chiral operators with $h_\mu < q_\mu/2$, in many interesting theories these do not show up in
 the above OPE.   Taking the limit $z\to 0$, defines the topological heterotic ring---a finite sub-ring of the $\cQb$-cohomology.

The existence of this structure implies that the genus zero correlators 
\begin{equation*}
c_{\alpha\beta\gamma} = \la \cO_{\alpha}(z_1) \cO_{\beta}(z_2) \cO_{\gamma}(z_3)\ra
\end{equation*}
in the half-twisted theory must be independent of the world-sheet metric and in particular of the insertion points $z_{1,2,3}$.  Our goal is to compute these correlators
in (0,2) LG theories.  The first step is to identify the elements of the topological heterotic ring. This a simple matter: the structure of the $\cQb$-cohomology allows to identify representatives for chiral operators in the UV, and the ``off-shell'' left-moving algebra formed by $T,j$ may be used to identify 
operators satisfying $h=q/2$ and to show that the OPE of these operators does define a sensible ring structure.

\subsection{The Topological Heterotic Ring of (0,2) LG Models}

A look at the action of $\cQb$ in eqn.~\ref{eq:susy} shows that, up to addition of holomorphic derivatives, the most general 
$\cQb$-closed operator has the form
\begin{equation*}
 \sum_{m=0}^n \gammab_{A_1} \cdots \gammab_{A_k} \gamma^{B_1} \cdots \gamma^{B_l} \omega^{A_1\cdots A_k}_{B_1\cdots B_l,i_1\cdots i_m} \psib^{i_1} \cdots \psib^{i_m},
\end{equation*}
where the coefficients $\omega(\phi,\phib)$ are constrained by
\begin{equation*}
\left[ -\psib^i \frac{\p}{\p\phib^i} +J_A \frac{\p}{\p\gammab_A} \right] \cO[\omega] = 0.
\end{equation*}
It is not hard to see that up to $\cQb$-exact terms the sum on $m$ collapses to the $m=0$ term
\begin{equation}
\label{eq:ops}
\cO[\omega] = \gammab_{A_1} \cdots \gammab_{A_k} \gamma^{B_1} \cdots \gamma^{B_l} \omega^{A_1\cdots A_k}_{B_1\cdots B_l}(\phi),
\end{equation}
where $\omega(\phi)$ is anti-symmetric in $A_{1},\ldots, A_k$ and in $B_1,\ldots,B_l$ and satisfies
\begin{equation}
\label{eq:omegacond}
J_{A_1} \omega^{A_1 \cdots A_k}_{B_1\cdots B_l} = 0.
\end{equation}
The operator $\cO[\omega]$ is $\cQb$-exact if and only if 
\begin{equation}
\label{eq:omegaexact}
\omega^{A_1\cdots A_k}_{B_1\cdots B_k} = J_A \eta^{A A_1 \cdots A_k}_{B_1\cdots B_l}
\end{equation}
for some $\eta$.

Since $\cO[\omega]$ is in $\cQb$-cohomology, we can reliably compute its 
$\GUL$ charge as well as weight with respect to $T$ by computing its OPE with $j$ and $T$.  Without loss of generality, we may assume $\cO[\omega]$ has a definite charge $q[\cO]$ under the $\GUL$ symmetry.
Picking any non-vanishing summand, say with indices $A_\alpha$, $\alpha=1,\ldots,k$
and $B_{\beta}$ with $\beta =1,\ldots, l$, we have
\begin{eqnarray*}
q[\cO] &=& \sum_{\alpha} (-Q_{A_\alpha}) + \sum_\beta Q_\beta + q [\omega] , \nonumber\\
h[\cO] &=& \frac{1}{2} \sum_{\alpha} (-Q_{A_\alpha}) + \frac{1}{2} \sum_\beta(1+Q_{B_\beta}) + \frac{1}{2} q[\omega],
\end{eqnarray*}
from which we conclude that $\cO$ has a definite weight under $T$, given by
\begin{equation}
h[\cO] = l + \frac{1}{2} q[\cO].
\end{equation}
This gives the desired bound $h \ge q/2$ for the chiral ring elements, which immediately implies that the LG theory
has a well-defined topological heterotic ring spanned by $\cO$ with $l=0$.\footnote{This conclusion is not altered
by the presence of chiral operators with extra holomorphic derivatives, since these contribute positively to $h(\cO)$.}   
In what follows, we will only consider $\omega$ with $l=0$.

\subsection{A Mathematical Interpretation}
The $\cO[\omega]$ have a mathematical interpretation in terms of the Koszul complex $K_\bullet$ associated to the ideal $J\subset R$~\cite{Kawai:1994qy}.  The Koszul complex is a standard tool in commutative algebra.  It is well-described in~\cite{Eisenbud:1995ca,Bruns:1993cm}, and many of its applications to residues and algebraic geometry are covered in~\cite{Griffiths:1978pa}. Here we will only need its most basic features.  To describe $K_\bullet$, let $\cE= R^N$ and think of $J = (J_1,\ldots,J_N)$ as an element in the dual module $\cE^\vee$.  The complex $K_\bullet$ is 
\begin{equation}
K_\bullet =
\begin{array}{l}
\xymatrix{0 \ar[r] &\wedge^N \cE \ar[r]^-{J\lrcorner} & \wedge^{N-1} \cE  \ar[r] &\cdots  \ar[r]  &\wedge^{k+1} \cE \ar[r]^-{J\lrcorner} &\wedge^{k} \cE \ar[r] &\cdots}\\
~~~~~~~~~~~~~\xymatrix{\cdots \ar[r] & \wedge^{2} \cE \ar[r]^-{J\lrcorner} & \cE \ar[r]^-{J\lrcorner} &R \ar[r] & 0 ,}
\end{array}
\end{equation}
where the map $J \lrcorner$ is just the interior product:
\begin{equation}
\begin{array}{lclcl}
J\lrcorner &:& \wedge^{k+1}\cE &        \to& \wedge^k\cE \\
J\lrcorner &:&     \omega^{AB_1\cdots B_k} &\mapsto&  J_A \omega^{A B_1\cdots B_k}.
\end{array}
\end{equation}
We denote the homology groups of this complex by $H_k(K_\bullet,J\lrcorner)$.  Eqns.~(\ref{eq:omegacond},\ref{eq:omegaexact}) show that the operators $\cO[\omega]$ with $\omega \in \wedge^k \cE$ are in one-to-one correspondence with elements of 
$H_k(K_\bullet,J\lrcorner)$.  

We will compute these groups in a couple of examples below, but for now we will just review a few simple properties.  The most basic property is 
\begin{equation*}
H_0(K_\bullet,J\lrcorner) = R/J.
\end{equation*}   
A less obvious property of the Koszul complex for an ideal $J$ in an arbitrary Cohen-Macaulay ring $R$ is that it provides a measure of the dimension of $J$.  Namely,
\begin{equation*}
H_{k}(K_\bullet, J\lrcorner) = \left\{\begin{array}{lll} 0 & \text{for} & k > N- \codim(J) \\ \text{non-zero} & \text{for} & k= N-\codim(J). \end{array} \right.
\end{equation*}
Without venturing into the depths and grades of dimension
theory, we state the result relevant for our zero-dimensional ideal $J$ in $R = \C[\phi_1,\cdots,\phi_n]$:  $\codim(J) = n$, so that
\begin{equation*}
H_k(K_\bullet, J\lrcorner) = 0\quad \text{for}\quad k > N-n,\quad \text{and}\quad H_{N-n}(K_\bullet,J\lrcorner) \neq 0.
\end{equation*}

The $H_k$ have a nice product structure:  given $\omega_1 \in H_{k_1}$ and $\omega_2 \in H_{k_2}$, we see that $\omega_1\wedge \omega_2 \in H_{k_1+k_2}$.  We expect this multiplicative structure to show up in the topological heterotic ring as well. 

We may equivalently work with the dual Koszul complex $K^\bullet$ and its cohomology groups $H^k(K^\bullet,J\wedge)$.  The entries in $K^\bullet$ are just $\wedge^k \cE^\vee$, and the maps are given by 
\begin{equation}
\begin{array}{lclcl}
J\wedge &:& \wedge^k\cE^{\vee} &        \to& \wedge^{k+1}\cE^{\vee} \\
J\wedge &:&     \omegat_{B_2\cdots B_{k+1}}      &\mapsto&  (k+1) J_{[B_1} \omegat_{ B_2\cdots B_{k+1}]}.
\end{array}
\end{equation}
The groups $H_k(K_\bullet, J\lrcorner)$ and $H^{N-k}(K^\bullet, J\wedge)$ are isomorphic, with isomorphism given by
the anti-symmetric $\ep$-tensor:
\begin{equation*}
\begin{array}{lclcl}
\ep &:&  H_{k}(K_\bullet, J\lrcorner) &        \to& H^{N-k}(K^\bullet, J\wedge) \\
\ep &:&   \omega^{A_1\cdots A_k}         &\mapsto&  \omega^{A_1\cdots A_k} \ep_{A_1 \cdots A_k B_1 \cdots B_{N-k}}.
\end{array}
\end{equation*}
We will have use for this isomorphism in our examination of the half-twisted correlators.  We will show that in addition to the multiplicative structure on the $H_k(K_\bullet, J\lrcorner)$, the correlators also yield a map $H^n (K^\bullet,J\wedge) \to \C$.

\subsection{The Singular Locus}
In most cases of interest the  $J_A$ depend on some parameters that cannot be absorbed into field re-definitions.  Such parameters should correspond to marginal deformations of the SCFT, and various physical quantities depend on the parameter values.  For small changes in the parameters we expect a smooth variation in the physical quantities, but for sufficiently large deformations the theory may become singular.  Essentially the only way a singularity can arise is due to appearance of new supersymmetric vacua.  Since the $J_A$ are quasi-homogeneous, additional vacua lead to a non-compact vacuum moduli space.  This non-compactness is the source of the singularity.  The sub-variety in parameter space where the dimension of the vacuum moduli space jumps is the singular locus.

The Koszul homology groups provide an algebraic criterion for the LG theory to be smooth:  all $H_k$ with $k>N-n$ must be trivial.  The singular locus may contain a number of irreducible components, and a
generic point on the singular locus should correspond to a situation where $H_{N-n+1} \neq 0$,
while points where different components meet may lead to non-vanishing $H_k$ with $k > N-n+1$.

When $N=n$, there is a more familiar criterion for the LG theory to be non-singular:  $\Delta = \det_{i,j} J_{i,j}$ must 
satisfy $\Delta \neq 0$ in R/J~\cite{Lerche:1989uy}.  This does not seem to be well-known when $J_i \neq \p W/\p \phi^i$, but a relatively simple proof of the assertion may be found in theorem 3.7 of~\cite{Tsikh:2004rc}, where it is shown that $\Delta \neq 0$ in $R/J$ if and only if J is zero-dimensional.

%One direction of the theorem is easy to see by using the Koszul homology groups.  Suppose $\omega \in \wedge^k \cE$ satisfies $J\lrcorner \omega = 0$.  Differentiating with respect to $\phi^i$, we find $J_{,i} \lrcorner \omega = 0$ in $R/J$.  Multiplying by the co-factor matrix for $J_{i,j}$ leads to $\Delta \omega = 0$ in $R/J$.  Thus, either $\Delta = 0$ or $\omega = 0$ in $R/J$.  In the latter case it is easy to show that $\omega$ must be given by $\omega = J\lrcorner\eta$.  We conclude that whenever $H_k(K_\bullet,J\lrcorner) \neq 0$ for $k>0$, then 
%$\Delta = 0$ in R/J.

\subsection{Comparison with (2,2) LG}
Before we turn to the correlators, we will review some standard (2,2) notions in the language above.
In (2,2) theories, $N=n$ and $J_i = \p W/\p \phi^i$.  The only non-vanishing Koszul homology group is $H_0 \simeq R/J$, so
that the observables are just elements of $R/J$, i.e., the usual (2,2) LG chiral ring.  Choosing a basis $\{\omega_\alpha\}$ for
$R/J$, all properties of the chiral ring are determined by the three-point functions 
$\la \cO[\omega_\alpha] \cO[\omega_\beta] \cO[ \omega_\gamma]\ra$ in the B-twisted topological field theory.
The result takes the form of the local Grothendieck residue~\cite{Vafa:1990mu}:
\begin{equation}
\label{eq:grotres}
\la \cO[\omega_\alpha] \cO[\omega_\beta] \cO[ \omega_\gamma]\ra = \frac{1}{(2\pi i)^n} \int_\Gamma \frac{\omega_\alpha\omega_\beta\omega_\gamma~ d\phi^1\wedge \cdots\wedge d\phi^n}{J_1 \cdots J_n},
\end{equation}
where $\Gamma$ is a cycle 
\begin{equation*}
\Gamma = \{ \phi~~|~~ | J_i |^2 = \ep_i > 0\}
\end{equation*}
oriented by $d \arg J_1 \wedge \cdots \wedge d\arg J_n \ge 0$.
The correlators are independent of choice of representatives, since the integral is invariant under $\omega_\alpha \to \omega_\alpha + f^i_\alpha J_i $.  The result is also independent of the $\ep_i$~\cite{Griffiths:1978pa}.  For simple examples this form often suffices, but for more complicated examples one must turn to
algebraic techniques to compute the residue~\cite{Kunz:2008rd}.

In~\cite{Melnikov:2007xi} it was shown that under (0,2)-preserving deformations of the (2,2) LG theory (i.e., $J_i\neq \p W/\p\phi^i$) the observables continue to be elements of $R/J$, and the correlators are computed exactly by the residue formula above.  In what follows
we will generalize these results to the much richer case of (0,2) models without a (2,2) locus.

%%%%%%%%%%%%%%%%%%%%%%%%%%%%%%%%%%%%%%%%%%%%%%%
\section{Localization of the Half-Twisted Lagrangian} \label{s:loc}
We will now construct the half-twisted action and use a localization argument to reduce the computation of correlators
to an integration over the zero-modes of the kinetic operators.  This idea is well-known in the case of (2,2) theories and the 
associated topologically twisted theories~\cite{Witten:1991zz}, and has recently been applied to the chiral ring of a number of (0,2) theories---see, e.g.,~\cite{Guffin:2007mp,Guffin:2008pi,McOrist:2008ji}.

\subsection{The Half-Twist}
The first step is to alter the spins of the fermion fields, shifting them by the $\GUB$ symmetry.  That is, given the 
Lorentz generator $J_T$, under which $\gamma, \gammab$ have charge $-1/2$, while $\psi,\psib$ have charge $+1/2$, we 
define a new Lorentz generator $J_{T'} = J_T - J_B/2$.  The modified Lorentz charges are collected in table~\ref{tab:charges}.
\begin{table}[ht]
\begin{center}
\begin{tabular}{|c|c|c|c|c|c|}
\hline
$~			$&$\phi^i		$&$\gamma^A	$&$\gammab_A	$&$\psi^i 	 	 $&$\psib^i		$\\ \hline
$J_{T'}		$&$0			$&$-1		$&$0			 	$&$+1		 $&$ 0			$\\ \hline
\end{tabular}
\end{center}
\caption{The twisted Lorentz charges}
\label{tab:charges}
\end{table}
To reflect the twist, we rename the fermionic fields according to
\begin{eqnarray}
\gamma^A &\to& \eta^A_z \nonumber\\
\gammab_A &\to& \chi_A \nonumber\\
\psi^i		   &\to& \rho^i_{\zb} \nonumber\\
\psib^i	   &\to & \theta^i.
\end{eqnarray}
Most importantly, under this twist the supercharge $\cQb$ becomes a world-sheet scalar BRST operator.  Thus, the half-twisted
theory naturally computes correlators of operators in the $\cQb$ cohomology---these are just the operators in the (0,2) chiral ring.  A formal argument shows that the correlators are independent under variations of the BRST-exact terms in the action.  This argument may fail if the vacuum manifold is non-compact, but we expect the independence to hold in the LG models. 
Letting $\cQ_T = -\cQb /\sqrt{2}$, the non-trivial (anti)-commutators are
\begin{equation}
\ACO{\cQ_T}{\phib^i} = -\theta^i,~~~\ACO{\cQ_T}{\rho^i_{\zb}} = 2\pb \phi^i,~~~\ACO{\cQ_T}{\chi_A} = J_A.
\end{equation}
The half-twisted action may be obtained from the untwisted action of eqn.~\ref{eq:Lcomp} by simply renaming the fermionic
fields.  We find it convenient to consider a one-parameter family of actions $S_t$, with the original action obtained at $t=1$:
\begin{equation}
\label{eq:Stwist}
S_t = \frac{1}{4\pi} \int d^2 z \left\{ 2 \eta^A_z \pb \chi_A - \eta^A_z J_{A,i} \rho^i_{\zb} + \ACO{\cQ_T}{2t\rho^i_{\zb} \p \phib^i + \chi_A \Jb^A}\right\}.
\end{equation}
Since $\cQ_T$-exact operators decouple in correlators of $\cQ_T$-closed operators, the theory must be invariant under deformations
of the $\cQ_T$-exact terms in $S$, and we may vary $t$ and $\Jb$ without affecting the correlators. We wish to compute the correlators via the path integral
\begin{equation}
\cN \int D[\text{fields}]  \cO[\omega_1](z_1) \cO[\omega_2](z_2) \cO[\omega_3](z_3) e^{-S_t},
\end{equation}
where $\cN$ is a normalization constant.  We fix the world-sheet to be a sphere with volume $4\pi V$.  Since the correlators are topological,
we expect them to be independent of the volume.

\subsection{Localization}
The integral is easily managed once we recall that it localizes onto configurations annihilated by
$\cQ_T$~\cite{Witten:1991zz}.  This means we can compute the correlator by expanding the action to leading order in fluctuations around configurations annihilated by $\cQ_T$. In the LG theory, the fixed point of $\cQ_T$ consists of the  point $\phi^i = 0$. It is convenient to do a partial localization to $\phi^i = \text{constant}$ and expand the fluctuations in the normalized eigenmodes of the 
scalar Laplacian on the sphere labelled by their eigenvalues $\lambda$.  

Note that the measure has charge $n-N$ under the $\GUB$ 
symmetry and charge $-r$ under the $\GUL$ symmetry.  Hence, if $\omega_{\alpha}\in \wedge^{k_\alpha} \cE$, a non-zero correlator
requires
\begin{equation}
k_1 + k_2 + k_3 = N-n,
\end{equation}
and
\begin{equation}
q_L[\cO [\omega_1] ]+q_L[\cO[\omega_2]]+q_L[\cO[\omega_3]] = r.
\end{equation}
A non-vanishing correlator then takes the form
\begin{equation}
\cN \int D[\text{fields}]_0 \chi_{A_1} \cdots \chi_{A_{N-n}} \omega^{A_1\cdots A_{N-n}}(\phi_0) \exp[-S_0] \times \prod_\lambda Z(\lambda),
\end{equation}
where $\omega = \omega_1\wedge \omega_2\wedge \omega_3$, 
\begin{equation}
S_0 = V \left[ J_A(\phi_0) \Jb^A(\phi_0) + \chi_A \Jb^A_i(\phi_0) \theta^i\right],
\end{equation}
and $Z(\lambda)$ is the contribution from the non-zero modes of the Laplacian with eigenvalue $\lambda$. $Z(\lambda)$ is given
by 
\begin{equation*}
Z(\lambda) = \int \prod_i d^2\phi^i  d\rho^id\theta^i \prod_A d \eta^A d\chi_A \exp[-S_\lambda],
\end{equation*}
where 
\begin{equation}
S_\lambda = t \lambda^2 \phi^i \phib^i + t \lambda \rho^i \theta^i - \lambda \chi_{A} \eta^A + \phi^i J_{A,i}(\phi_0) \Jb^{A}_j(\phi_0)\phib^j + \chi_A \Jb^A_i(\phi_0) \theta^i.
\end{equation}
This Gaussian integral is easy enough to do exactly, but the $t$-independence makes the evaluation especially simple.
Under a change of variables
\begin{equation*}
\phi \to t^{-1} \phi,~~~ \rho \to t^{-1} \rho,~~~\theta \to t^{-1} \theta
\end{equation*}
we find that the measure is invariant, while the action becomes
\begin{equation}
S_\lambda = \lambda^2 \phi^i \phib^i + \lambda \rho^i \theta^i-\lambda\chi_A\eta^A + \text{O}(t^{-1}).
\end{equation}
Taking $t\to\infty$ leads to $Z(\lambda) = \pi^{n} \lambda^{N-n}$.
In contrast to topologically twisted path integrals (or their (0,2) deformations), the half-twisted path integral
has a divergence that must be regulated.  In the case at hand this is a benign zero-point energy divergence that may be subtracted
in a parameter-independent fashion.  We choose to absorb it into the normalization constant $\cN$.

Finally, we are left with a finite-dimensional integral over the $\phi, \chi$ zero-modes.  Integrating out the $\chi_A$, we
are left with an integral over $\C^n$:
\begin{equation*}
\la \cO[\omega_1] \cO[\omega_2] \cO[\omega_3] \ra = \cN V^n \int \prod_i d^2\phi^i \ep_{A_1\cdots A_r B_1 \cdots B_n} \omega^{A_1\cdots A_r} \Jb^{B_1}_1\cdots \Jb^{B_n}_{n} e^{ - V J_C \Jb^C }.
\end{equation*}
We choose a convenient normalization and express the integrand in terms of {$\omegat \in H^{n}(K^\bullet,J\wedge)$}, the dual
of $\omega = \omega_1\wedge \omega_2 \wedge\omega_3$:
\begin{equation}
\label{eq:corint}
\la \cO[\omega_1] \cO[\omega_2] \cO[\omega_3] \ra  =  \int \prod_i \frac{ d^2\phi^i}{\pi} \omegat_{A_1\cdots A_n} \Jb^{A_1}_1\cdots \Jb^{A_n}_{n} e^{ - J_C \Jb^C }.
\end{equation}
Note that we have absorbed a factor of the world-sheet volume $V$ into the $\Jb_A$.  We will see momentarily that this is justified.

%%%%%%%%%%%%%%%%%%%%%%%%%%%%%%%%%%%%%%%%%%%%%%%
\section{Correlators and Residues} \label{s:res}
Eqn.~(\ref{eq:corint}) reduces the computation of the topological heterotic ring to the determination of the 
map $\tau: H^{n}(K^\bullet, J\wedge) \to \C$ given by
\begin{equation}
\label{eq:tauint}
\tau(\omegat) =  \int_{\C^n} d\mu ~\omegat_{A_1\cdots A_n} \Jb^{A_1}_1\cdots \Jb^{A_n}_{n} e^{-S},
\end{equation}
where $S = J_C \Jb^C$, and the measure $d\mu$ is 
\begin{equation*}
d\mu = \frac{i}{2\pi} d\phi^1 \wedge d\phib^1\wedge \cdots \wedge \frac{i}{2\pi} d\phi^n \wedge d\phib^n.
\end{equation*}
Although the integral is well-behaved, its form hides some important features and is not particularly suited to explicit computations.  In this section we will improve on this state of affairs.    

\subsection{Some Properties of $\tau$}
We begin by posing two questions about the map $\tau$:
\begin{enumerate}
\item The map uses a specific representative of $\omegat \in H^{n}(K^\bullet, J\wedge)$.  Does $\tau$ depend on the 
         choice of representative?
\item Suppose the $J_A$ depend on a parameter $\alpha$.  Does $\tau$ depend 
         on $\alphab$, as eqn.~(\ref{eq:tauint}) seems to suggest? 
\end{enumerate}
Each query has a negative answer, and in each case, the underlying reason is that $\cQ_T$-exact deformations of the action do
not affect the correlators of the half-twisted theory.  We will now verify these properties directly from eqn.~(\ref{eq:tauint}). 

To see that the map is independent of representatives, observe that
\begin{eqnarray}
\tau (J\wedge \etat) &=& - \int_{\C^n} d\mu~\frac{ \ep^{i_1\cdots i_n}}{(n-1)!} \etat_{A_2\cdots A_n} \frac{\p}{\p\phib^{i_1}}\left[J_{A_1} \Jb^{A_1}\right] \Jb^{A_2}_{i_2}\cdots\Jb^{A_n}_{i_n} e^{-S}
\nonumber\\
~ & = & ~\int_{\C^n} d\mu \frac{\p}{\p\phib^{i_1}} \left[-  \frac{ \ep^{i_1\cdots i_n}}{(n-1)!} \etat_{A_2\cdots A_n} \Jb^{A_2}_{i_2}\cdots\Jb^{A_n}_{i_n} e^{-S}\right].
\end{eqnarray}
The integrand is a total derivative, and the factor of $e^{-S}$ ensures that there are no dangerous boundary terms.  Thus, $\tau(J\wedge \etat) = 0$.

To determine the dependence of $\tau$ on $\Jb$ we consider a small variation $\delta \Jb$ and compute the corresponding 
change in $\tau$.  We find
\begin{eqnarray}
\delta \tau (\omegat) & = & \int_{\C^n} d\mu~\omegat_{A_1\cdots A_n} \left[\delta(\Jb^{A_1}_1 \cdots \Jb^{A_n}_n) - \Jb^{A_1}_1 \cdots \Jb^{A_n}_n J_B \delta\Jb^B \right] e^{-S}.
\end{eqnarray}
Since $J \wedge \omegat = 0$, it follows that
\begin{equation}
\omegat_{A_1 \cdots A_n} J_B = n ~\omegat_{B [A_2 \cdots A_n} J_{A_1]}.
\end{equation}
Use of this identity in the second term in $\delta \tau(\omegat)$ leads to
\begin{eqnarray}
\delta \tau (\omegat) & = & \int_{\C^n} d\mu~ \frac{\p}{\p\phib^{i_1}}\left[ \frac{ \ep^{i_1\cdots i_n}}{(n-1)!} \omegat_{A_1\cdots A_n} \delta\Jb^{A_1} \Jb^{A_2}_{i_2} \cdots \Jb^{A_n}_{i_n} e^{-S} \right].
\end{eqnarray}
We conclude that $\delta \tau = 0$, so that $\tau$ is ``independent of the choice of $\Jb^A$.''  To make that precise, consider a one-parameter family $\Jb^A_\lambda$ for $\lambda \in [0,1]$, with $\Jb^A_1 = \overline{J_A(\phi)}$.  We have shown that $\tau_0 = \tau_1$, provided the integral defining $\tau_\lambda$ converges for all $\lambda \in [0,1]$.  So, for any $V\neq 0$ rescaling $\Jb^A \to V \Jb^A$ as was done above does not affect $\tau$.  In addition, we see that $\tau$ has a holomorphic dependence on parameters in the $J_A$.  

A slight modification of these two arguments also shows that $\tau(f \omegat) = 0$ for any $f \in J$ and $\omegat \in H^n(K,J\wedge)$:
\begin{eqnarray}
\tau (\omegat J_B) & = & \int_{\C^n} d\mu~ n \omegat_{B A_2\cdots A_n} J_{A_1} \Jb^{[A_1}_{1} \cdots \Jb^{A_n]}_n e^{-S}
\nonumber\\
~&= &  \int_{\C^n} d\mu \frac{\p}{\p\phib^{i_1}} \left[ -\frac{\ep^{i_1\cdots i_n}}{(n-1)!} \omegat_{BA_2\cdots A_n}\Jb^{[A_2}_{i_1} \cdots \Jb^{A_n]}_{i_n} e^{-S} \right]  =0.
\end{eqnarray}
It follows that $\tau(\omegat f) = 0$ for all $f\in J$.
This is consistent with the property that the ideal $J$ annihilates the Koszul cohomology groups.

\subsection{The Non-Degenerate Case and $N=n$}
Although the most interesting physical applications require $J$ to be 
quasi-homogeneous and maximally degenerate (i.e., $\rank(J_{A,i}) =0 $ at $\phi=0$), $\tau$ is actually well-defined for any
zero-dimensional ideal $J$.  When $J_{A,i}$ has maximal rank at every zero of the $J_A$, the integral may be computed exactly by a saddle-point approximation around the solutions to $J_A = 0$.  Labelling these by $\phi_p$, the result is
\begin{equation}
\label{eq:trsum}
\tau(\omegat) = 
\sum_{\phi_p} \left. \frac{ \omegat_{A_1\cdots A_n} \Jb^{A_1}_1 \cdots \Jb^{A_n}_n}{\det_{i,j} J_{B,i} \Jb^B_{j} } \right|_{\phi = \phi_p},
\end{equation}
We stress that, despite appearances, $\tau$ is independent of the $\Jb^A$.  

When $N=n$, the $\Jb$-independence becomes obvious. In this case, 
\begin{equation*}
\omegat^{A_1 \cdots A_n} = \ep^{A_1\cdots A_n} \times f,
\end{equation*}
for some $f \in R/J$, and the map takes the form
\begin{equation}
\tau(\omegat) = \sum_{\phi_p}   \frac{f(\phi_p)}{\cJ_J(\phi_p)},
\end{equation}
where $\cJ_J$ is the Jacobian
\begin{equation*}
\cJ_J(\phi) =\left| \frac{\p( J_1, \cdots ,J_n)}{\p(\phi_1,\cdots,\phi_n)} \right|.
\end{equation*}
This can be re-cast as a local Grothendieck residue of eqn.~(\ref{eq:grotres}):
\begin{equation*}
\tau(\omegat) = \frac{1}{(2\pi i)^n} \int_{\Gamma} \frac{f(\phi) ~~d\phi^1\wedge\cdots \wedge d\phi^n}{J_1\cdots J_n}.
\end{equation*}
The advantage of this form is that it remains sensible even in the degenerate case when $\cJ_J(\phi_p) = 0$.  By considering
relevant deformations of $J$, i.e., deformations that do not bring in new solutions to $J_A=0$ from infinity, one can
argue that even in the degenerate case (as long as $N=n$) $\tau$ is given by the Grothendieck residue~\cite{Vafa:1990mu,Melnikov:2007xi}.

\subsection{A Residue for $N>n$}
It would be nice if it were possible to express $\tau$ as a multi-variate residue when $N > n$.  We have
not been able to find such a result in all generality.  However, there is an important special case where there is a natural residue
formula.  Consider again the case of a quasi-homogeneous ideal $J$ with an isolated zero at the
origin.  Suppose that there exists a set $\sigma = \{B_1,B_2,\ldots,B_n\} \subseteq \{1,\ldots,N\}$ such that the ideal generated by
$K_i= J_{B_i}$  has an isolated zero at the origin.  Introduce a parameter $\lambda$ by replacing $\Jb_A \to \lambda \Jb_A$
whenever $A \not\in\sigma$.  In this case the map takes the form
\begin{equation}
\tau(\omegat) = \int_{\C^n} d\mu~ \left[\omegat_{B_1 B_2 \cdots B_n} \det_{i,j} \Kb^i_{j} + O(\lambda)\right] ~e^{-K_i \Kb^i +O(\lambda) } \quad\text{(no sum on the $B_i$)}.
\end{equation}
We have shown that $\tau$ is independent of small changes in $\lambda$, and we expect that we may safely take
$\lambda \to 0$ as long as the integral continues to converge.  But the condition on the ideal $K$ assures that the exponential factor
continues to provide the necessary convergence.  Thus, we may safely take $\lambda \to 0$ and reduce the integral to
the case when $N= n$.  Following the discussion of the $N=n$ case above, we are led to a residue formula for $\tau$:
\begin{equation}
\label{eq:taures}
\tau(\omegat) = \frac{1}{(2\pi i)^n} \int_\Gamma \frac{ \omegat_{B_1\cdots B_n} d\phi^1\wedge\cdots\wedge d\phi^n}{J_{B_1} \cdots J_{B_n}}, \quad\text{(no sum on the $B_i$)},
\end{equation}
where the cycle $\Gamma$ is given by
\begin{equation}
\Gamma = \{ \phi~|~ |J_{B_i}|^2 = \ep_i > 0\},
\end{equation}
with orientation fixed by $d \arg(J_{B_1}) \wedge \cdots\wedge d \arg(J_{B_n}) \ge 0$.
As a check that eqn.~(\ref{eq:taures}) is sensible, we note that it satisfies the basic expectations $\tau(J\wedge \etat) = 0$ and $\tau(f~\omegat ) = 0$ for all $\etat\in \wedge^k \cE^\vee)$, $f \in J$, and $\omegat \in H^n(K,J\wedge)$.

The residue form for $\tau$ is important.  It means that whenever there exists a subset $\sigma$ with the desired properties,
we may compute $\tau$, and therefore the topological heterotic ring of the (0,2) LG theory, by using the powerful algebraic
techniques developed for the study of residues.   In models of most immediate physical interest---where the LG theory describes a point
in the moduli space of a linear sigma model for a Calabi-Yau target-space constructed as a complete intersection in a toric variety---the 
subset $\sigma$ certainly exists.  More generally, the requisite $\sigma$ may not exist, as is illustrated by an example with $n=2,N=3$ and 
\begin{equation}
J_1 = \phi_1\phi_2^3,\quad J_2 = \phi_2(\phi_1^2-\phi_2),\quad J_3 = \phi_1 (\phi_1^2-\phi_2).
\end{equation}
Nevertheless, the original form for $\tau$ in eqn.~(\ref{eq:tauint}) still satisfies all the nice properties expected from a residue, and it may well be that the theory of residue currents may be of use in 
unravelling its algebraic structure.

\section{Examples} \label{s:examples}
\subsection{An $n=1$ case}
As we saw above, models with $n=1$ are not interesting,  since in the IR they are equivalent to a (2,2)-supersymmetric $A_{k}$ 
minimal model tensored with a free theory of left-moving Weyl fermions.  Nevertheless, the case deserves a look because $\tau$ may be computed by using any of its forms. To be concrete, we pick $N=2$ with $J_1 = J_2 = \phi^{k+1}$.  In this
case, the Koszul homology groups are
\begin{eqnarray}
H_0 (K_\bullet, J\lrcorner) &=& R/J = \{1, \phi, \ldots,\phi^k\}, \nonumber\\
H_1 (K_\bullet, J\lrcorner) &=& \{(\phi^m,-\phi^m)^T\}, \quad m=0,\ldots,k.
\end{eqnarray}
The three-point functions have the form
\begin{equation}
\la \cO[\omega] \cO[\phi^{l_1}] \cO[\phi^{l_2}]\ra = \tau( \omegat),
\end{equation}
where $\omegat = (\phi^l,\phi^l)$ and $l = m+l_1+l_2$.
Using eqn.~(\ref{eq:tauint}), we have
\begin{eqnarray}
\tau(\omegat) & = & \int \frac{d^2\phi}{\pi} \omegat_A \p_{\phib} \Jb^A e^{-2 \phi^k \phib^k} \nonumber\\
~ & = &~ \delta_{l,k} \int_0^\infty dy~ 2 (k+1) y^k e^{-2 y^{k+1}} = \delta_{l,k}~~,
\end{eqnarray}
where in the second line we substituted $y = \phi \phib$ and integrated over the angular coordinate.

Alternatively, we may use the residue form in eqn.~(\ref{eq:taures}) with the subset $\sigma = \{1\}$ or $\sigma=\{2\}$.
In either case,  the cycle $\Gamma$ is a counter-clockwise contour surrounding $\phi = 0$ and
\begin{equation}
\tau(\omegat) =\int_\Gamma \frac{ d\phi}{2\pi i} \frac{ \phi^l}{\phi^{k+1}} = \delta_{l,k}.
\end{equation}

\subsection{An N=3,n=2 Example}
We choose
\begin{equation}
J_1 = \phi_1^4 + \phi_1^2\phi_2^3,\quad J_2 = \phi_2^4, \quad J_3 = \phi_1^5.
\end{equation}
This ideal is quasi-homogeneous, and the natural normalization  for the charges given in eqn.~(\ref{eq:chargenorm}) leads to
\begin{equation}
q_1 = \frac{3}{14},\quad q_2 = \frac{1}{7},\quad Q_1 = -\frac{6}{7},\quad Q_2  = -\frac{4}{7},\quad Q_3 = -\frac{15}{14}.
\end{equation}
The expected $\GUL$ normalization and central charges are then given by
\begin{equation}
r = \frac{15}{7},\quad c = \frac{31}{7}, \quad \cb = \frac{24}{7}.
\end{equation}
There are two non-trivial Koszul cohomology groups, $H^3(K^\bullet,J\wedge) \simeq R/J$, and $H^2(K^\bullet,J\wedge)$.  To determine
the structure of $H^2(K^\bullet,J\wedge)$, we seek solutions to
\begin{equation}
\omegat_{12} J_3 + \omegat_{31} J_2 + \omegat_{23} J_1 =  0.
\end{equation}
A simple elimination computation shows that $\omegat$ may be written in terms of two elements $\alpha,\beta \in R$:
\begin{equation}
\omegat_{12} = \alpha\phi_1-\beta\phi_2,\quad\omegat_{31} = - \beta\phi_1^3 -\alpha\phi_1^2\phi_2^2,\quad\omegat_{23} = \alpha(\phi_2^3-\phi_1^2)+\beta\phi_1\phi_2.
\end{equation}
The pair $\alpha,\beta$ yield a trivial element in $H^2(K^\bullet,J\wedge)$ if and only if
\begin{eqnarray}
\alpha & = &\etat_1 \phi_1\phi_2 -\etat_2 \phi_1^3 - \etat_3 \phi_2, \nonumber\\
\beta  & = & \etat_1(\phi_1^2-\phi_2^3) + \etat_2 \phi_1^2\phi_2^2 - \etat_3 \phi_1
\end{eqnarray}
for some $\etat\in \cE^\vee$

We now wish to determine $\tau$.  There are two ways to express 
$\tau$ in residue form:  we may either take $\sigma = \{1,2\}$ or $\sigma = \{2,3\}$.  In the first case, we find
\begin{equation}
\tau(\omegat) = \int_{\Gamma} \frac{d\phi_1\wedge d\phi_2}{(2\pi i)^2} \frac{\omegat_{12}}{J_1 J_2},
\end{equation}
where
\begin{equation}
\Gamma = \{~\phi~|~ |J_1|^2 = \ep_1^8,\quad |J_2|^2 = \ep_2^8 \}.
\end{equation}
Since the residue is independent of the parameters $\ep_1,\ep_2$, we can choose these so that the cycle
$\Gamma$ takes a simple form.  A convenient limit is to take $\ep_1^{-2}\ep_2^3 \ll 1$.  We parametrize
$\phi_2$ by $\phi_2 = \ep_2 e^{i\theta_2}$, $0\le \theta_2 \le 2\pi$, and let $\phi_1 = \ep_1 u$, so that 
the condition $|J_1|^2 =\ep_1^8$ takes the form
\begin{equation}
|u^4 + \delta u^2|^2 = 1, 
\end{equation}
where $\delta = \ep_1^{-2} \ep_2^3 e^{3i\theta_2}$.  When $|\delta| \ll 1$, $|J_1|^2 =\ep_1^8$ describes a $\phi_2$-dependent
contour in the $u$-plane:
\begin{equation}
u = e^{i\theta_1} -\frac{1}{4} \delta e^{-i\theta_1} + O(\delta^2),\quad 0\le \theta_1\le 2\pi.
\end{equation}
Up to small corrections this is just a circle of radius $\ep_1$ in the $\phi_1$ plane which encloses the zeroes of $J_1$.
The orientation of $\Gamma$ is determined by $d\theta_1 \wedge d\theta_2 \ge 0$, so that 
\begin{equation}
\tau(\omegat) = \int_{\Gamma_2(\ep_2)} \frac{d\phi_2}{2\pi i} \int_{\Gamma_1(\ep_1,\phi_2)} \frac{d\phi_1}{2\pi i} ~~\frac{\omegat_{12}}{\phi_2^4(\phi_1^4+\phi_1^2\phi_2^3)}.
\end{equation}
In principle, the integration must be carried out  in the order indicated; however, since the integrand is holomoprhic and $\Gamma_1$ has only a weak dependence on $\phi_2$, we may replace $\Gamma_1$  by $\phi_1 = \ep_1 e^{i\theta_1}$.  Now we may safely exchange the
order of integration and compute the integral by residues. The result is
\begin{equation}
\tau(\omegat) = - \frac{1}{24} \p_1^4 \alpha + \frac{1}{12} \p_1^2 \p_2^3 \alpha - \frac{1}{12} \p_1^3 \p_2^2 \beta .
\end{equation}
This answer is consistent with the $\GUB$ and $\GUL$ selection rules, and it is easy to check that 
$\tau(J\wedge \etat ) = \tau(f \omegat) = 0$ for all $\etat \in \cE^\vee$ and for all $f \in J$.

Next, we take $\sigma = \{2,3\}$.  In this case, the cycle $\Gamma$ is described as $\Gamma_1\times \Gamma_2$, with $\Gamma_1 : |\phi_1|^2=\ep_1^2$ and $\Gamma_2 : |\phi_2|^2= \ep_2^2$, oriented by $d\theta_2 \wedge d\theta_1 \ge 0$.  Thus,
\begin{equation}
\tau(\omegat) = - \int_{\Gamma_1} \frac{d\phi_1}{2\pi i} \int_{\Gamma_2} \frac{d\phi_2}{2\pi i} ~\frac{\omegat_{23}}{\phi_1^5\phi_2^4}.
\end{equation}
Computing this residue we recover the answer obtained from $\sigma = \{1,2\}$.

%%%%%%%%%%%%%%%%%%%%%%%%%%%%%%%%%%%%%%%%%%%%%%%

%\bibliographystyle{utphys}
%\bibliography{bigref}
%\bibliographystyle{/Users/lmel/BIB/utphys}
%\bibliography{/Users/lmel/BIB/bigref}
\providecommand{\href}[2]{#2}\begingroup\raggedright\endgroup

\end{document}